\documentclass[acmsmall]{acmart}

\pdfoutput=1
\usepackage{amsmath,amsfonts}
\usepackage{algorithmic}
\usepackage{graphicx}
\usepackage{enumitem}
\usepackage{textcomp}
\usepackage{listings}

\usepackage{longtable}
\usepackage{float}
\usepackage{graphicx}
\usepackage{multirow}
\usepackage{balance}
\usepackage{hyperref}
\usepackage{tabularx} 
\usepackage{booktabs} 
\usepackage{colortbl}
\usepackage{hhline}
\usepackage{array}
\newcolumntype{L}[1]{>{\raggedright\let\newline\\\arraybackslash\hspace{0pt}}m{#1}}
\newcolumntype{C}[1]{>{\centering\let\newline\\\arraybackslash\hspace{0pt}}m{#1}}
\newcolumntype{R}[1]{>{\raggedleft\let\newline\\\arraybackslash\hspace{0pt}}m{#1}}
\usepackage{subfigure}
\usepackage{enumitem}

\newenvironment{boxedtext}
    {
    
    \begin{center}

    \begin{tabular}{|p{0.96\linewidth}|}
    \hline
    }
    { 
    \\ \hline
    \end{tabular} 
    
    \end{center}
       }

\newcommand{\subsc}[1]{\vspace{3pt} \noindent  {\textbf{\emph{#1}}}}

\definecolor{dartmouthgreen}{rgb}{0.05, 0.5, 0.06}

\AtBeginDocument{%
  \providecommand\BibTeX{{%
    \normalfont B\kern-0.5em{\scshape i\kern-0.25em b}\kern-0.8em\TeX}}}

\setcopyright{cc}
\setcctype{by-nc}

\begin{document}
\title{The Role of Social Identity in Shaping Biases Against Minorities in Software Organizations}

\author{Sayma Sultana}
\email{ssultana1@tulane.edu}
\affiliation{%
  \institution{Tulane University}
  \city{New Orleans}
  \state{Louisiana}
  \country{USA}
}

\author{London A. Cavaletto}
\email{london.cavaletto@wayne.edu}
\affiliation{%
  \institution{Wayne State University}
  \city{Detroit}
  \state{Michigan}
  \country{USA}
}

\author{Bianca Trinkenreich}
\email{bianca.trinkenreich@colostate.edu}
\affiliation{%
  \institution{Colorado State University}
  \city{Fort Collins}
  \state{Colorado}
  \country{USA}
}

\author{Amiangshu Bosu}
\email{amiangshu.bosu@wayne.edu}
\affiliation{%
  \institution{Wayne State University}
  \city{Detroit}
  \state{Michigan}
  \country{USA}
}

\renewcommand{\shortauthors}{Sultana \em{et} al.}

\begin{abstract}
While systemic workplace bias is well-documented in non-computing fields, its specific impact on software engineers remains poorly understood. This study addresses that gap by applying Social Identity Theory (SIT) to investigate four distinct forms of bias: lack of career development, stereotyped task selection, unwelcoming environments, and identity attacks. Using a vignette-based survey, we quantified the prevalence of these biases, identified the demographics most affected, assessed their consequences, and explored the motivations behind biased actions.
Our results show that career development and task selection biases are the most prevalent forms, with over two-thirds of victims experiencing them multiple times. Women were more than three times as likely as men to face career development bias, task selection bias, and an unwelcoming environment. In parallel, individuals from marginalized ethnic backgrounds were disproportionately targeted by identity attacks. Our analysis also confirms that, beyond gender and race, factors such as age, years of experience, organization size, and geographic location are significant predictors of bias victimization.


\end{abstract}




\maketitle

\definecolor{codegreen}{rgb}{0,0.6,0}
\definecolor{codegray}{rgb}{0.5,0.5,0.5}
\definecolor{codepurple}{rgb}{0.58,0,0.82}
\definecolor{backcolour}{rgb}{0.95,0.95,0.92}

\lstdefinestyle{mystyle}{
    backgroundcolor=\color{backcolour},   
    commentstyle=\color{codegreen},
    keywordstyle=\color{magenta},
    numberstyle=\tiny\color{codegray},
    stringstyle=\color{codepurple},
    basicstyle=\ttfamily\footnotesize,
    breakatwhitespace=false,         
    breaklines=true,                 
    captionpos=b,                    
    keepspaces=true,                 
    numbers=left,                    
    numbersep=5pt,                  
    showspaces=false,                
    showstringspaces=false,
    showtabs=false,                  
    tabsize=2
}

\lstset{style=mystyle}

\newcommand{\surveyquote}[1]{\begin{addmargin}[1em]
{0em}\emph{#1 }\end{addmargin} \vspace{2pt}}

\newcommand{\etal}{{\emph{et} al. }}

\definecolor{coolblack}{rgb}{0.0, 0.18, 0.59}
\newcommand{\excerpt}[1]{\textcolor{coolblack}{``\emph{#1}''}}

\definecolor{background_gray}{gray}{0.85}
\definecolor{background_green}{rgb}{7,163,90}

\newcommand{\bestvalue}[1]{\cellcolor{background_gray}\textbf{#1}}

\newcommand{\greenshade}[1]{\cellcolor{background_green}\textbf{#1}}
\definecolor{MidnightBlue}{rgb}{0.1, 0.1, 0.44}

\newcommand{\code}[1]{{\tt #1}}

\newcommand{\newkeyword}[1]{ \textcolor{blue}{ #1}}

\newcommand{\significant}[1]{ \cellcolor{gray!25} #1}

{\small \textcolor{red}{ Warning: This paper contains examples of language that some
people may find offensive or upsetting.} }
\section{Introduction}
\label{sec:sec-introduction}

Diverse teams, by drawing insights from a variety of backgrounds~\cite{vasilescu2015gender,herring2009does}, are critical for developing a deeper understanding of and effective response to the needs of the user bases for whom software is designed~\cite{wagner2006creation,zallio2022designing}. Such teams are more likely to anticipate the impacts of technical solutions on distinct user groups, thereby yielding inclusive, accessible, and ethically sound software~\cite{szlavi2023gender,easterbrook2008selecting}. However, the current landscape of software development organizations remains characterized by a distinct lack of diversity~\cite{ashcraft2016women,stackoverflow_survey_2022}. This underrepresentation of marginalized groups not only limits creative problem-solving capabilities and productivity~\cite{vasilescu2015gender,lebovits2019automating,buolamwini2018gender}, but also fosters an echo chamber of ideas. Such homogeneity may inadvertently perpetuate biases in software systems and reinforce broader societal inequalities~\cite{wang2020reducing,wang2019implicit,galhotra2017fairness}.

Despite industry efforts to promote the recruitment and retention of underrepresented groups~\cite{guizani2022perceptions,mozilla-diversity,google-diversity}, research indicates the persistence of numerous implicit and explicit biases~\cite{sultana2023code,bosu2019diversity,breidenbach2021implicit,wang2019implicit,lee_carver} that hinder the creation of an equitable environment. While recent studies have begun to investigate prevalent biases~\cite{rodriguez2021perceived,trinkenreich2022empirical,guzman2024mind}, significant gaps remain. First, while literature from non-computing domains has established that demographic characteristics drive disparities in opportunities and treatment~\cite{bertrand2004emily}, the specific association between software engineers'\footnote{Hereinafter, we use `software engineers' as an umbrella term to refer to various software development roles such as developers, testers, analysts, database administrators, architects, maintainers, etc.} demographic profiles, their susceptibility to biases, and their subsequent responses remains scantily explored. Second, effective interventions require an understanding of the actors exhibiting biased behavior; however, investigations targeting the perspective of the perpetrator within software engineering are currently missing. Therefore, this study aims to \textit{understand how a software engineer's social identity (e.g., demographics) shapes their susceptibility to biases and the resultant consequences.}

We investigate these biases through the lens of Social Identity Theory (SIT)~\cite{tajfel1979integrative}, which posits that individuals derive a sense of identity and self-worth from their social group memberships. Biases frequently emerge during interactions between in-groups (those perceived as belonging) and out-groups (those perceived as different), often reinforcing in-group dominance while limiting support and career growth for out-group members~\cite{chow2004gender}. The application of SIT is particularly relevant given the historical evolution of the computing domain. In its formative years, programming roles were strongly shaped by social perceptions regarding who was “fit” for technical work~\cite{abbate2012recoding}. As the profession expanded, dominant groups consolidated power through professional associations and selection mechanisms that often disadvantaged those outside the majority~\cite{ensmenger2012computer}. Over time, the community embraced narrow, exclusionary stereotypes of the “ideal” programmer~\cite{ensmenger2012computer}, solidifying boundaries between in-groups and out-groups. Furthermore, leadership roles in contemporary organizations remain disproportionately concentrated among dominant social groups~\cite{bosu2019diversity,abbate2012recoding}, perpetuating dynamics that influence access to opportunities. Conversely, the establishment of minority-oriented forums within Free and Open Source Software (FOSS) projects~\cite{debian-lists,women-moz} illustrates how individuals organize around shared identity to resist exclusion. Because SIT explains how group-based categorizations influence behavior and offers insights into interventions to counteract bias~\cite{khadka2024social}, it serves as the foundational framework for this study.

\vspace{2pt} \noindent \textbf{Approach:} Guided by the SIT framework, we focus on four specific categories of bias: i) stereotyped task selection~\cite{treude2023she}, ii) lack of career development opportunities~\cite{trinkenreich2022empirical}, iii) identity attacks~\cite{margolis2002unlocking}, and iv) unwelcoming environments~\cite{thomae2015sexist}. We designed an online survey utilizing vignettes—hypothetical scenarios depicting social dilemmas—to elicit participants' experiences. For each vignette, respondents indicated the frequency with which they had encountered or witnessed such incidents and described the impact on the involved parties. The survey was disseminated through social media channels and minority-oriented developer mailing lists. Following the application of six exclusion criteria, the final dataset comprised 253 valid responses. We employed regression analysis to determine the influence of respondent demographics on the observation or experience of bias. Additionally, we conducted a thematic analysis to examine the impact of these incidents on victims and to characterize the motivations of those who actively or passively contributed to the biased behaviors.

\vspace{2pt}
\noindent \textbf{Key findings:}  Career development and task selection biases are the most prevalent, with over two-thirds of victims experiencing biases multiple times. Beyond gender, factors such as age, years of development experience, project exposure, race, and location significantly influenced the likelihood of bias victimization.
Women were more than three times as likely as men to experience career development bias, task selection bias, and an unwelcoming environment. Meanwhile, individuals from marginalized ethnic backgrounds were disproportionately targeted by identity attacks.
Additionally, workplace culture and stereotypical beliefs about women often influence individuals to take biased actions.

\vspace{2pt}
\noindent \textbf{Primary contributions}:
\begin{itemize}
    \item A comparative analysis of four types of biases among software development organizations.      
    \item An empirical investigation of consequences arising from biases and their associations with demographic factors.
    \item An investigation of the motivation behind the people who actively or passively contribute to biased actions.
    \item Actionable insights to identify and mitigate biases in computing organizations.
    \end{itemize}

The remainder of this paper is organized as follows. 
Section~\ref{sec:research-questions} presents the three research questions investigated in this study.
Section~\ref{sec:related-works} briefly overviews closely related works.
Section \ref{sec:research-method} details our research methodology. 
Section \ref{sec:results} presents the results of our analysis.
Section \ref{sec:discussion} discusses key implications based on our findings. Finally, Section \ref{sec:threats} and \ref{conclusion} present the threats of validity and conclude the paper, respectively.

\section{Research Questions}
\label{sec:research-questions}
The primary goal of this study is to \textit{understand how a software engineer's social identity shapes his/her susceptibility to biases and the resultant consequences.} We aim to achieve this goal with the following three research questions.


\vspace{2pt}
\noindent \textbf{(RQ1) How do various demographic factors associate with being victims of biases?}

\noindent \textit{Motivation:}  
Prior works indicate that biases against individuals with multiple marginalized identities do not operate in isolation. Rather, distinct systems of inequality, such as sexism, racism, and caste discrimination, intersect and exacerbate one another~\cite{solanke2009putting,connor2023intersectional}. Joan Williams characterizes this compounding effect as ``double jeopardy,'' highlighting that individuals may simultaneously encounter multiple forms of bias that mutually reinforce one another~\cite{williams2014double}. She further emphasizes that the manifestation of a specific bias is often modulated by intersecting identities, including race, ethnicity, immigration status, socioeconomic background, and place of origin~\cite{williams2014double}.

While existing Software Engineering (SE) literature has extensively examined gender biases~\cite{imtiaz2019investigating,trinkenreich2022empirical,sultana2023code}, the impact of broader social identities, such as age, ethnicity, geography, and education, remains largely understudied. A notable exception is the work of Nadri et al., which found that non-white developers face higher rejection rates for pull requests~\cite{nadri2021relationship}. Consequently, a systematic investigation into these correlations is warranted. Expanding this scope is essential to fully understand how various identities influence a developer's susceptibility to bias and to identify the specific subgroups most vulnerable to ``double jeopardy.''

\vspace{2pt}
\noindent \textbf{(RQ2) What are the consequences of biases on victims and people around them?}

\noindent \textit{Motivation:} 
Prior Software Engineering (SE) studies have cataloged various biases encountered by practitioners~\cite{Bias_in_STEM,gender_influence_2,treude2023she, ellemers2004underrepresentation,thomae2015sexist,trinkenreich2022empirical}. However, limited empirical insight exists regarding the downstream effects of these biases on victims and their professional environments. Responses to bias are multifaceted; a victim may internalize the bias as a professional norm, actively resist it, alter their career trajectory, or reluctantly develop coping mechanisms, often suffering short- or long-term psychological distress~\cite{parker2002so}.

Furthermore, the specific reaction of a victim is often modulated by demographic factors, which are deeply entwined with cultural norms and expectations~\cite{heilman1983sex}. Consequently, a robust understanding of these consequences, and their intersectionality with demographic characteristics, is critical for developing context-specific mitigation strategies.


\vspace{2pt}
\noindent \textbf{(RQ3) What are the motives of the biased actors?}

\noindent 
\textit{Motivation:}  
Prior works have illuminated the motivations of biased actors through a variety of theoretical lenses. For instance, the ``lack of fit'' model has been used to explain why specific demographic groups are systematically overlooked for managerial positions~\cite{heilman1983sex}, while frameworks of benevolent sexism interpret the assignment of simpler tasks to marginalized groups not as hostility, but as a form of patronizing protection~\cite{kuchynka}. Furthermore, Social Identity Theory provides a mechanism for understanding stereotyping and identity attacks as functions of in-group and out-group dynamics~\cite{reid2010language,Brylla2023}, and workplace harassment has been framed as a manifestation of broader systems of power and organizational hierarchy~\cite{stockdale1993sexual}.

Despite this theoretical breadth, the firsthand perspectives of the software engineers who act as the sources of these biases remain conspicuously absent from the literature. Existing theories largely observe behavior from the outside or focus on the victim's experience. Consequently, they do not account for how the unique technical pressures, meritocratic ideals, or collaborative workflows of software engineering might specifically incentivize or rationalize biased behavior. Investigating the perpetrators' own articulated motivations, whether conscious or unconscious, is essential. Without understanding these root causes, mitigation strategies risk being superficial. Meaningful intervention requires addressing the specific cognitive and cultural drivers of bias within the engineering workforce.
\section{Related Works}
\label{sec:related-works}


\textbf{Forms of biases against minorities in computing:} 
Women are significantly underrepresented in technology, holding just 25\% of computing occupations. While our study examined multiple demographic dimensions, gender-based experiences emerged as the most prominent theme among prior studies investigating biases in computing. 
Prior studies confirm that biases systematically place women at a disadvantage. 
Stereotypes and structural obstacles like the "glass ceiling" and "maternal wall" limit women's access to leadership roles and key technical tasks \cite{maji2020gendered,misa-dynamics-2021,treude2023she,canedo2021breaking,trinkenreich2022empirical,shantha2024breaking}.
Harassment, hostility, and identity attacks create unwelcoming workplaces, particularly for women and LGBTQ+ individuals \cite{margolis2002unlocking,oliveira2024navigating,soares2023investigating}. Biases also manifest as pay gaps, hiring discrimination, and a general distrust of women's technical abilities \cite{segovia2020being,campero2021hiring,weststar2018women,oliveira2024navigating,trinkenreich2022empirical}. Subtle forces like benevolent sexism and tokenism also reinforce traditional roles and undermine genuine inclusion \cite{trinkenreich2022empirical,kaushik2018study}.
Together, these factors perpetuate exclusion and underscore the urgent need for systemic change.

\vspace{2pt}
\noindent \textbf{Biases in FOSS communities:} While diversity and inclusion in Free and Open Source Software (FOSS) have gained attention, underrepresentation remains a persistent issue across gender, race, ethnicity, age, and geography~\cite{bosu2019diversity,nadri2021relationship,rastogi2016biases,wachs2022geography}. A significant body of research has moved beyond documenting this imbalance to investigate how systemic biases manifest in core FOSS activities and shape contributor experiences.
Evidence of differential treatment is prominent in the evaluation of contributions. Studies have shown that a contributor's perceived identity can have a direct influence on outcomes. For instance, Nadri \textit{et al.} found that developers perceived as `White' have their pull requests accepted at a higher rate than non-White developers~\cite{nadri2021relationship}. Similarly, geography and nationality create bias, with contribution acceptance being influenced by a developer's location~\cite{rastogi2016biases,wachs2022geography}. This scrutiny extends to the code review process, where a contributor's race, ethnicity, and age can affect the degree of pushback they receive on their work~\cite{murphy2022pushback}.
Beyond direct contribution evaluations, structural and social barriers disproportionately affect newcomers and those from non-dominant groups. Steinmacher \textit{et al.} identified social and onboarding hurdles, such as cultural differences and language barriers, that can prevent successful integration into a project~\cite{steinmacher2019overcoming}.
The consequence of these systemic issues is often a negative personal experience that discourages long-term participation. Contributors from marginalized groups frequently limit their engagement due to concerns about their competence, a lack of belonging, or the anticipation of negative treatment~\cite{wang2018competence,sultana2023code}. In many cases, these fears are realized through direct experiences with stereotyping, undervaluation, aggression, and harassment~\cite{lee_carver,singh2021motivated}. Collectively, prior works highlight that biases in FOSS are not isolated incidents but are deeply embedded in community interactions and workflows, thereby reinforcing patterns of underrepresentation.

\vspace{2pt}
\noindent \textbf{Biases in computing industry:} 
{Despite broader participation compared to FOSS projects, biases continue to persist in commercial software development~\cite{bosu2019diversity,blincoe2019perceptions,griffiths2010disappearing}. Wang and Redmiles’ Implicit Association Test (IAT) revealed unconscious biases influencing hiring decisions and contribution evaluations~\cite{wang2019implicit}. Blincoe \textit{et al.} noted 40\% of survey respondents admitted unfair treatment, yet gender discrimination was unreported in interviews, hinting at hidden struggles. With 82\% of women in IT feeling unheard~\cite{raja2022discovering}, exclusion and unconscious bias hinder their growth~\cite{griffiths2010disappearing, hyrynsalmi2019underrepresentation}. Exclusion and unconscious bias have been shown to limit opportunities for growth and recognition across multiple identity groups~\cite{griffiths2010disappearing,hyrynsalmi2019underrepresentation}. In addition, explicit forms of bias, such as resistance during code reviews~\cite{murphy2022pushback}, bullying, and harassment, further contribute to the marginalization and underrepresentation of certain populations in software engineering~\cite{bandias2009women}. Researchers have also examined how different forms of bias intersect in computing, including racism~\cite{twine2018technology,twine2022geek}, ageism~\cite{sharma2016workplace,hyrynsalmi2019motivates,bandias2009women,griffiths2010disappearing}, and veteran status~\cite{van2023still}. Twine, for example, documented how employees in Silicon Valley experienced overlapping racial and gender biases, with Black and Latin women encountering compounded disadvantages~\cite{twine2018technology,twine2022geek}. Similarly, age has been shown to influence employment status, career advancement, and income levels in the ICT sector~\cite{bandias2009women}. These findings highlight that workplace inequalities in computing are not shaped by a single dimension of identity, but rather by the intersection of multiple identity factors that amplify or alter individuals’ experiences of bias.}

\vspace{2pt}
\noindent \textbf{Intersectionality with other forms of biases:} Prior research has also extended the analysis of biases against minorities to its intersections with other systems of inequality, including racism~\cite{twine2018technology, twine2022geek}, ageism~\cite{sharma2016workplace,hyrynsalmi2019motivates, bandias2009women, griffiths2010disappearing}, and veteran status~\cite{van2023still} within the computing domain. Specifically, Twine documented that Black and Latina women in Silicon Valley technology firms navigate a distinct intersection of racism and sexism~\cite{twine2018technology, twine2022geek}. These findings indicate that women from minoritized ethnic groups frequently face employment discrimination despite possessing the requisite technical proficiencies.

Furthermore, Adya et al. revealed that the lived experience of women in computing is distinctively shaped by race. Their work found that American women perceived higher levels of gender stereotyping and salary disparities compared to their South Asian counterparts~\cite{adya2008women}. Similarly, age acts as a critical compounding factor. For example, research indicates that age significantly influences women's employment status, career advancement, and income levels within the Australian ICT sector~\cite{bandias2009women}.

\vspace{2pt}
\noindent \textbf{Novelty:} Despite existing research on biases in computing, critical questions remain unanswered. We investigate: (RQ1) the effect of various demographic factors on software engineers' susceptibility to biases from the lens of Social Identity; (RQ2) how the consequences of biases vary across different demographic groups; and (RQ3) the motivations behind biased actions by software engineers.

\section{Research Methodology}
\label{sec:research-method}
\begin{figure*}
	\centering  \includegraphics[width=0.95\linewidth]{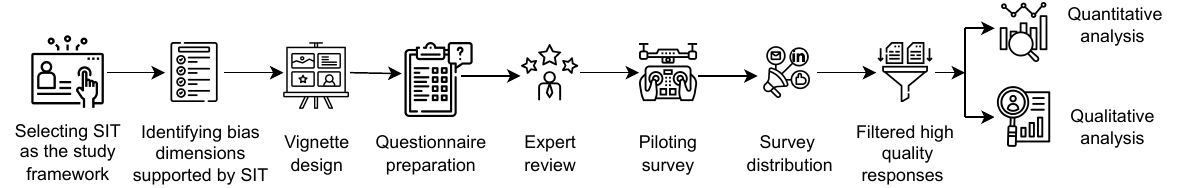}
  \caption{An overview of our research method}
\label{fig:research method}	
\end{figure*}

Figure~\ref{fig:research method} shows an overview of our research methodology, detailed in the following subsections.

\begin{table}
    \centering
    \caption{Survey questionnaire}
    \label{tab:survey_questions}

\resizebox{\linewidth}{!}{

\begin{tabular}{p{1.5cm}|p{1cm}|p{1cm}|p{6cm}|p{4cm}}
\rowcolor[HTML]{9B9B9B} 
\hline
Group & \#Q & RQ & Question  Text   & Response Option    \\
\hline

\rowcolor[HTML]{EFEFEF} 
\multirow{5}{*}{Receiver} & Q4.3 & RQ1 & Reflecting on my career in the software industry, I have experienced a similar scenario to the one experienced by Melissa (victim) as described in the story.  &    \textit{Yes; May be; No} \\ 

 & Q4.4 & RQ1 &  Following the previous question, where did you have this experience? Check all that apply. &  At my current organization; At a previous organization; At multiple organizations   \\ 
 \rowcolor[HTML]{EFEFEF}  
& Q4.5 & RQ3 & How many times did you have this experience?                                        & Once; Twice; 3-5 times; 5-10 times; More than 10 times;   \\
                                                                                 
& Q4.6 & RQ3 & Think back to the time when you experienced the scenario first-hand, what can you recall from it, how did you feel during the incident, and what was the impact of such a scenario on you? Share what you feel most comfortable with.           & Open-ended question      \\ 

\rowcolor[HTML]{EFEFEF} 
& Q4.7 & RQ3 & After the scenario took place, how were you and the other participants affected? Share what you feel most comfortable with.                                &  Open-ended question    \\

\hline

 & Q4.8 & RQ1 & Reflecting on my career in the software industry, I have witnessed/heard of a same/similar scenario happen, but I was not a participant.                                     & Yes; May be; No  
 \\

\rowcolor[HTML]{EFEFEF} 
\multirow{4}{*}{Witness} & Q4.9 & RQ1 & How many times did you witness/hear of the same/similar scenario take place?  & Once; Twice; 3-5 times; 5-10 times; More than 10 times;  \\                                                                                                    
& Q4.10 & RQ3 & Think back to the time when you witnessed/heard of the scenario take place, what can you recall from it?    & Open-ended   \\ 

\rowcolor[HTML]{EFEFEF} 
& Q4.11 & RQ3 & After the scenario took place, how were the participants affected?        &  Open-ended \\ 
\hline

\multirow{3}{1.5cm}{Biased actors} & Q4.12 & RQ1 & I can think of at least one time when I have made the same/similar decision as Sean (the perpetrator) about a potential candidate. &      Yes; May be; No       \\ 

\rowcolor[HTML]{EFEFEF} 
& Q4.13 & RQ1 & Following the previous question, how often did you make such a decision? & Once; Twice; 3-5 times; 5-10 times; More than 10 times; \\ 

& Q4.14 & RQ4 & Think back to the time where you made the same/similar decision as Sean, who were you making the decision about, what was the decision, and why did you make that decision? & Open-ended \\
\hline

\end{tabular}      
}

\end{table}

\subsection{Biases through the lens of SIT}
SIT, developed by Tajfel and Turner~\cite{tajfel1979integrative}, explains how group-based biases form in social settings. SIT is built on three core concepts: individuals derive part of their self-concept from membership in social groups, which leads to a natural preference for their own "in-group" (in-group favoritism), and a corresponding bias against those in the "out-group" (out-group discrimination).
In a technical workplace, these dynamics can lead to tangible inequities. For example, in a demographically homogenous software team, a developer from an underrepresented group may be unconsciously perceived as an "outsider." This perception can lead the dominant in-group to undervalue the outsider's competence, regardless of his/her skills and qualifications.
Consequently, in-group favoritism may manifest as specific, observable biases, such as reserving complex, high-visibility tasks for one another and recommending each other for promotions, or sharing knowledge within exclusive informal networks.

While SIT explains a broad spectrum of biases, our study investigates four specific forms for two primary reasons.
First, the four biases we selected are prominently and consistently linked to the SIT framework in prior research on workplace dynamics. This provides a strong theoretical foundation for our investigation.
Second, this focus is a deliberate methodological choice to ensure the collection of high-quality data. Research on survey design emphasizes that keeping completion time under 20 minutes is critical for maintaining participant engagement and data reliability~\cite{revilla2017ideal}. Including a more extensive list of biases would have created a lengthy survey, potentially resulting in lower completion rates and compromising the validity of the responses.
Therefore, our investigation is strategically centered on the following four well-defined biases.

\vspace{2pt}
\noindent \textbf{1. Career Development opportunities Bias (CDB)}
Social Identity Theory (SIT) posits that individuals favor their in-group over out-group members~\cite{tajfel1979integrative}, creating workplace barriers for minorities. This bias limits access to networks, mentorship, and promotions, contributing to a ``glass ceiling'' effect~\cite{meijers1998development,soto2021senior}. It also drives gender pay gaps, discriminatory hiring, lack of recognition, and the ``maternal wall'' and leadership disparities~\cite{chow2004gender}.

\vspace{2pt}
\noindent \textbf{2. Stereotyped Task Selection Bias (TSB)}
Social Identity Theory (SIT) suggests that people categorize themselves and others into groups to simplify social contexts~\cite{tajfel1979integrative}, often leading to stereotyping. In-group members receive positive traits, while out-groups face negative assumptions~\cite{reid2010language}, such as minorities being seen as less competent. This impacts professional settings, with women assigned ``pink tasks'' (high-effort, low-visibility) in team projects~\cite{garcia2022gender,brough2011women} and biased task associations in software development (e.g., eliciting requirements associated with women 94\% of the time)~\cite{treude2023she}. Segregated hiring practices also steer women into lower-status roles~\cite{campero2021hiring}. Stereotyping may also lead to benevolent sexism, ``prove-it-again'' bias, and tokenism~\cite{chow2004gender}.

\vspace{2pt}
\noindent \textbf{3. Unwelcoming Environment (UE)} 
According to SIT, members of minority groups may encounter an unwelcoming environment as a result of being perceived as different or “other” by dominant in-groups, which often leads to microaggressions or other forms of discrimination~\cite{maass2003sexual}. 
Moreover, patriarchal norms contribute to the development of a sexually aggressive belief system in specific individuals to a greater extent than in others, and this belief system, in turn, can lead to a tendency to misperceive interactions~\cite{stockdale1993sexual}. As a result, men often interpret their flirtations or sexual interests as harmless instead of harassment due to their misconception of women's intentions and reactions~\cite{margolis2002unlocking,elephant}.

\vspace{2pt}
\noindent \textbf{4. Identity attack (IA)} 
According to SIT, identity attacks focus on the fundamental elements of an individual's social identity, as in-group members may seek to delegitimize or devalue the identities of out-group ones~\cite{tajfel1979integrative}. Furthermore, when individuals have a strong identification with their group, any perceived threat to the group's status or value can be interpreted as a personal attack~\cite{Brylla2023}. This perception can lead to derogatory remarks or exclusionary behaviors directed towards out-group members.
Prior studies have found jokes and remarks targeting another person's sexual orientation, gender identity, religion, or ethnicity~\cite{thomae2015sexist,plester2009crossing,santos2023transgender}. Such incidents create an unwelcoming environment for the victims and may lead to poor performance and attrition.

\subsection{Questionnaire Design}
We designed one vignette for each of the four selected bias types. We added questions to inquire if a respondent had encountered/heard of / witnessed/ created such scenarios for others. If a respondent replies affirmatively, we follow up with in-depth questions to learn more about such incidents. We also included ten demographic questions to learn a respondent's background and enable an analysis of how demography contributes to experiencing biases in computing organizations. 

\subsubsection{Vignettes design}
Asking participants directly about biases, by naming or defining them, can be confusing and may lead to underreporting, as individuals may not always recognize specific incidents as biased~\cite{bell2021can}.
To facilitate recall and encourage honest reflection, we employed vignettes. 
Traditionally, vignettes serve to create psychological distance, separating the respondent from the vignette character~\cite{steiner2016designing}. However, we employed vignettes as a tool to facilitate recall, aligning with recent studies  that leverage them to help respondents recognize similar dynamics in their personal experiences, even when those experiences do not perfectly match the presented scenarios~\cite{goldney2001mental,thorsteinsson2014mental,kim2015influence}. This approach offers a nuanced balance between indirect probing and insightful reflection, thereby fostering more profound engagement with biased situations.

Prior software engineering studies also used vignettes to study destructive criticism in code review~\cite{gunawardena2022destructive} and investigate students' ethical decision-making~\cite {mcnamara2018does}. To make the vignettes credible and authentic to the participants, two were based on real-world incidents reported in a prior study~\cite{dunham2017third}, one based on a post on  devRant\footnote{https://devrant.com/}, and the remaining one based on an incident shared by the Everyday Sexism project\footnote{https://everydaysexism.com}. We also include contexts by mentioning the designations and short background of the characters. Though we have designed the scenarios in such a way that bias-based decisions or behavior exist, we did not mention the keyword `bias' to avoid prejudice. Each vignette in our survey: i) is relatively short (within 150 words), ii) includes at least two personas, where one persona represents a victim and the other represents the biased actor, and iii) interactions between the personas demonstrate biased behavior. After presenting a vignette, we specified the goal of the vignette to help respondents understand the behavior to focus on. For example, after introducing the first vignette on CDB, we asked, ``Can you think of a time when a similar scenario to the one described above occurred in an organization you have worked at or are currently working at? For example, \textit{a well-deserving candidate, like Melissa, did not receive a promotion.}"

\begin{enumerate}[leftmargin=*]

\item \textbf{Scenario 1: Career development bias (CDB).} We developed this scenario based on the Walmart vs Betty Dukes case, where Betty filed a lawsuit against Walmart for favoring men during promotion\cite{dunham2017third}. Despite having a strong work record, she was overlooked for promotion again and again. In our survey, we focus on the `lack of career development opportunities' for the persona Melissa, who is a hard-working woman. Her boss Sean (M), selects Asish, another man, for a promotion to a managerial position despite of Melissa being both senior and more qualified than Asish. This scenario depicts that women are passed over for promotion or challenging projects due to gender stereotyping.
On the contrary, Asish does not have such responsibilities for his family; therefore, he can focus on work more than Melissa. This scenario depicts that women are passed over for promotion or challenging projects since leadership roles are often considered to require masculine traits. 
    
\item \textbf{Scenario 2: Task selection bias (TSB).}  Kuchynka et al\cite{kuchynka2018hostile, kuchynka} shares that in STEM environments both hostile and benevolent sexism can take place by treating women as if they can not handle the challenge in STEM tasks  or they lack relative STEM competence. We found an exemplary scenario of benevolent sexism where women are not assigned challenging tasks due to the idea that men should support and take care of women\cite{sattari2022dismantling}. We develop the task selection bias scenario based on this example- there is a man persona named Eric, a woman, Gianna. Eric assigns low-complexity tasks, e.g., documentation and frontend development, to Gianna. Rarely is Gianna assigned more involved tasks, and at that time, she is closely monitored by Eric. When she encounters problems, instead of helping her solve them, Eric passes that task to Steve, a man who was hired at the same time as Gianna. 


\item \textbf{Scenario 3: Unwelcoming Environment (UE). } This scenario is designed from Pao vs Kleiner case\cite{dunham2017third} where Ellen Pao, a junior partner in a capital firm encountered sexual advances from her colleague and despite complaining to authorities did not receive any support. In the `unwanted sexual attention' scenario, the persona named Shruti, who is a woman, receives excessive attention from Kevin (a man), who works on the same project as Shruti. Despite being turned down by Shruti multiple times, Kevin keeps asking her out. She hesitates to file a formal complaint against him because that might not be considered harassment by company policy. So,  Shruti unofficially complains about Kevin to her supervisor,  Steve (a man), and he laughs it away, saying that he is harmless and she has nothing to worry about. 


\item \textbf{Scenario 4: Identity attack (IA). } In this scenario involving `identity attack', the persona named Mikehl is a gay man. Mikehl brings his partner, Christian, to the company's anniversary party and introduces  Christian to his colleagues. After that, Salvatore ( a straight man), Mikehl's colleague, starts making crude remarks targeting the sexual orientation of Mikehl during conversations and in code reviews. This vignette is designed being motivated by a post shared in devRant\footnote{https://devrant.com/rants/421271/what-do-you-call-a-gay-programmer-a-backend-engineer}.

\end{enumerate}

\subsubsection{Questions on a vignette} For each vignette, we ask if a respondent had encountered such a scenario in any organization he/she worked in. If they respond negatively, the subsequent questions on the vignette are skipped, and the respondents are moved to the next vignette. If the answer is affirmative, we ask them questions divided into three groups: 
    i) \emph{Receiver:} The respondent experienced the scenario as the receiver of the bias;
    ii) \emph{Witness:} The respondent was the witness to a biased scenario;
    iii) \emph{Bias actor:} The respondent has taken a biased decision or showed a biased attitude in that scenario.
When a respondent responds affirmatively regarding a group (e.g., as a recipient), we ask further questions about the incidents, such as how often they have encountered such an incident. On the other hand, for a negative response, we skip such questions. Each group includes closed and open-ended questions about the occurrence frequency and effect. To understand the prevalence of biased incidents, we asked the respondents where they experienced bias in the receiver question group. They can choose multiple answers to these questions. We have also asked how often they have experienced bias as receivers, witnesses, or biased actors. Furthermore, if the respondents belonged to the receiver or witness group, we asked them about their experiences and how those experiences affected them and the people around them. We kept these questions open-ended to gather as much information about the biased incidents in real-world settings and their influence as possible.  If the respondents belong to the biased actor group, we asked about their motivation behind the decision or attitude. Since these are sensitive topics, more in-depth questions might remind their traumatizing/ uncomfortable experiences. Therefore,  we asked them to share what they were comfortable with. All the questions were optional, and our questions were designed to put the respondents at ease. 

\subsubsection{Demographic questions} To characterize our respondents, our survey starts with ten demographic questions on gender, age group, location,  and years of experience,  as detailed in Table~\ref{tab:demography}. Since there is no common standard for studying ethnicity, we followed prior studies in software engineering domain for creating the list of ethnicity~\cite{weeraddana2023empirical,nadri2021relationship}. Our anonymous survey encourages our respondents to answer honestly without any risk of being identified. We did not ask/record any personally identifiable information, and we solicit all demographic data using range-based multiple-choice questions. 

\subsubsection{Survey validation}
We validated our survey using two steps. In the first step, we sent our questionnaire to three renowned software engineering researchers with extensive experience in survey research. Based on their suggestions, we shortened the scenarios, revised some of the question texts, added clarifications, and changed the options for the demography questions to comply with US Census Bureau guidelines~\cite{pan2013development}.  After these changes, we conducted a survey pilot with five CS graduate students who are diverse regarding gender, age, and ethnicity, each with multiple years of software industry experience. 
This pilot aimed to estimate survey completion time, identify comprehension issues, and obtain respondents' suggestions. We also tested the survey on multiple devices and platforms. This pilot helped to identify and fix several minor presentation issues and ensure the survey was accessible on multiple platforms.
Our survey instruments, recruitment strategy, and solicitation materials were reviewed and approved by our Institutional Review Board. We have included the complete survey questionnaire in the supplementary materials.

\subsection{Data collection}
 We collected survey responses in two phases. In the first phase, we advertised our Qualtrics-hosted survey on social media such as Facebook, Twitter, and LinkedIn. We requested participation from software engineers, who have worked in the industry within the last three years. We noticed high-quality responses during the first few weeks as we monitored response quality. However, we saw an increasing ratio of low-quality and potentially spam responses as time progressed. Since this issue is common for online anonymous surveys advertised on social media~\cite{teitcher2015detecting}, we closed the first phase after six weeks. We started the second phase with a different survey URL. Spam responses, especially if distributed through social media, are common~\cite{spam_weed_out}. Usually, these spammers have groups. Once a survey’s link is posted there, most of the responses will be spam. Weeding out spam is time-consuming. Hence, we created a new URL and closed the previous one. The questions are identical; only the links themselves are different. In this second phase, we sent invitations to various minority-oriented (e.g., women, African-American, LGBTQ+,  Latino, etc.)  software professionals' groups and mailing lists. Since our survey was conducted solely in English, we acknowledge the limitation of primarily posting on forums that predominantly use English. The second phase spanned eight weeks. 
 

\subsection{Data analysis}

We received 579 responses in two phases, where the first phase contributed 223. We established strict quality assessment criteria to exclude low-quality responses.
    \textcircled{\raisebox{-0.9pt}{1}} We excluded 244 responses for being incomplete, as indicated by Qualtrics.
    \textcircled{\raisebox{-0.9pt}{2}} We also discarded 22 responses where participants took less than 4 minutes to complete the survey since reading four vignettes and the minimum number of questions would require at least that interval. 
    \textcircled{\raisebox{-0.9pt}{3}} Qualtrics reported fraud score or duplicate score as more than 50\% for 18 of the responses. We excluded those.
    \textcircled{\raisebox{-0.9pt}{4}} We discarded 31 responses that included random garbage texts for open-ended questions.
    \textcircled{\raisebox{-0.9pt}{5}} We found irrelevant answers for the open-ended questions in 11 responses, suggesting a lack of understanding or spamming. Therefore, excluded. 
    
    Using these five criteria, we were left with 253 responses, 78 of which came from the first phase.  We conducted qualitative and quantitative data analysis to answer the proposed research questions.  We provided \$10 Amazon gift cards to 10 randomly selected respondents from the ones surviving our exclusion criteria.

\vspace{2pt}
\noindent \textbf{Qualitative data analysis} For each scenario, we collected responses detailing the repercussions of biased attitudes and the motives behind the actions. Our approach involved an inductive coding process to analyze the data. Initially, researchers familiarized themselves with the text responses through multiple readings. These readings also helped us diligently apply our exclusion criteria. Subsequently, they individually coded 100 responses,  25 from each bias type from the answers on consequences. Similarly, they coded all  33 responses from the answers on biased actors' motives. They had a discussion session to merge the codes with similar ideas and build an agreed-upon list of codes. This process created a list of 48 codes for RQ2 and 11 codes for RQ3. 
With these agreed-upon codes, two authors independently labeled each response related to RQ2 and RQ3 at the code level.  We measure the level of agreement among the labelers using Cohen's kappa, which suggests a substantial agreement  ($\kappa$ =0.65).
Consequently, the labelers had a discussion session to resolve all the conflicts through mutual agreements.  Finally, we had another discussion session to categorize the 48 codes for RQ2 (i.e., bias consequences) into six higher-level categories. Table~\ref{tab:qualitative_result} shows the list of codes and their assigned categories. We skipped this categorization step for RQ3 as very few respondents acknowledged being biased actors.

\vspace{2pt}
\noindent  \textbf{Quantitative data analysis} We use logistic regression models to examine the associations between demographic factors and the likelihood of victims of a bias type and how a person responds to such incidents. Since very few people acknowledged their role as biased actors, we skipped this analysis for RQ4 due to inadequate sample size to obtain statistically significant results. 
We trained an inferential logistic regression model for each bias type to estimate the association between demography and being a victim of that bias type. For example, the CDB model's dependent variable is \code{is\_victim\_CDB}, and variables representing demography are independent. 
For the categorical responses such as Race, Location, and Education, we used one-hot encoding, which converts each response choice for a demographic into a dichotomous value. However, we noticed some options, such as the number of respondents from South America, were selected very few times. Since the number of samples in our dataset is not large, a dichotomous variable \code{is\_location\_SA} will have limited influence on the model. To encounter this challenge, we aggregated multiple groups with small occurrences to a higher-level group. For example, respondents not located in North America or the European Union were merged into the group `Rest of the world.' Similarly, independent variables for gender are `men', `women', and `others'; for the highest degree, variables are `below-college', `college', and `post-grad';  and for ethnicity, variables are `White', `Asian', `African American', `Latino', and `Other races'.  We would like to mention that this limitation is not due to our survey design or data collection efforts. Despite our efforts to reach diverse groups of minorities, some groups remain scarce in our dataset, as they are among computing organizations.
We adopted the model construction and analysis approach from Harrell Jr.~\cite{harrell2015regression}. We assess model performance using Veall-Zimmermann pseudo-$R^2$~\cite{veall1994evaluating}. We also ran the log-likelihood test to compare each model against a null model to assess its goodness of fit. We assess each variable's significance and influence on the dependent using p-value and Odds ratio, respectively. 

\vspace{2pt}
\noindent \textbf{Validity analysis} We employed a convergent parallel mixed-methods design, using a single survey instrument to simultaneously collect both quantitative (e.g., scaled questions) and qualitative (e.g., open-ended responses) data~\cite{storey2024guiding}.
To ensure the interpretative validity of our results, we implemented two key strategies. First, following the convergent approach, we analyzed the quantitative and qualitative datasets separately, then integrated and synthesized the findings during the final interpretation. Second, we enhanced the rigor of our study through perspective triangulation by capturing the distinct viewpoints of individuals who have been victims of bias, witnesses to it, and actors who have committed it.

\section{Results}
\label{sec:results}
 The following subsections describe our respondents' demographics and detail the results of the three research questions. 
\begin{table}
    \caption{Demography of the survey respondent}
    \label{tab:demography}
       \centering

\resizebox{\linewidth}{!}{
\begin{tabular}{|p{2.5cm}|p{11cm}|}
\rowcolor[HTML]{9B9B9B} 
Demography                      & Group : \% Respondents     \\
Gender& Man: 36.8\%; Woman: 53.4\%; Nonbinary and Others: 7.1\%; Not disclosed: 2.3\%\\
\hline
Age & 18-20: 5\%; 21-25: 5.1\%; 26-30: 14.6\%; 31-35: 25.3\%; 36-45: 22.1\%; 46-55: 23.3\%;  56-65: 7.5\%; 66+: 2.0 \%\\
\hline
Software development experience & Less than a year: 4.3\%; Between one to five years: 34.4\%; Between six to ten years: 28.6\%; Between eleven to twenty years: 20.2\%; More than twenty years: 12.6\%\\
\hline
Tenure with current project /organization& Less than a year: 24.1\%; Between one to five years: 54.9\%; Between six to ten years: 13.8\%; Between eleven to twenty years: 6.7\%; More than twenty years: 0.4\%\\
\hline
Number of developers in current project /organization & 1-4: 12.1\%; 5-9: 10.2\%; 10-19: 12.9\%; 20-49: 13.4\%; 50-99 : 6.7\%; 100-249 : 12.2\%; 250-499 : 3.9\%;  500-999 : 9.0\%; 1000+ : 18.9\%\\
\hline
Is volunteer & Yes : 28.9\%; No : 71.1\%\\
\hline
Highest level of education& Less than high school: 1.9\%;  High school or equivalent: 5.9\%; Some college but no degree: 9.1\%; Associate degree: 3.9\%; Bachelor degree: 43.9\%;  Masters: 26.9\%; Ph.D.: 8.3\%\\ 
\hline
Current location & North America: 42.3\%; Europe: 34.4\%; South America: 2.8\%; East Asia : 7.1\%; Other Asian Countries: 8.7\%; Oceania: 2.8\%; Africa: 1.6\%\\
\hline
Race & White: 57.3\%; Asian: 26.5\%; Black / African American: 7.9\%; Hispanic or Latino: 5.5\%; 
American Indian / Alaska Native: 3.5\%; Hawaiian / Pacific Islander: 1.2\%; Others: 4.8\%\\
\hline
Project type & Commercial closed source:  41.7\%; Commercial open source: 16.6\%; Non-Profit open source: 33.2\%\\
\hline

\end{tabular}    

}
    \end{table}

\subsection{Respondent demographics}
Table~\ref{tab:demography} shows our respondents' demographics.
We observed that 135 (=53.4\%) of our respondents had identified themselves as women, 88 (=34.8\%) as men, and 18 (=7.1\%) respondents belonged to nonbinary or another group that was not listed. {Additionally, 67 respondents (=26.5\%) identified as Asian, 19 (=7.5\%) as African American, 14 (=5.5\%) as Hispanic or Latino, and 131 (=52\%) as White. Furthermore, our dataset included participants from diverse age groups and geography. The respondents' demographics align with our intention to gather mostly the experiences of minorities. Therefore, our respondents represent a population that is suitable for this study's goal.  

\begin{table}[]
    \caption{Frequency of the respondents experiencing Career Development Bias (CDB), Task Selection Bias (TSB), Unwelcoming Environment (UE), and identity attack (IA).}
        \centering     
    
    \begin{tabular}{p{6cm}p{1cm}p{1cm}p{1cm}p{1cm}}
    \hline
\textbf{\% of the respondents} & \multicolumn{1}{c|}{ \textbf{CDB (\%)}} &  \multicolumn{1}{c|}{\textbf{TSB (\%)}} & \multicolumn{1}{c|}{\textbf{UE (\%)}} & \multicolumn{1}{c|}{\textbf{IA (\%)}} \\ \hline
 Acknowledged existence of bias &36.4 &43.7 &30.7 &18.3\\ \hline 
 Been a victim &28.5 &33.2 &17.9 &10.1\\ \hline 
 Been a victim at current organization &14.2 &16.2 &3.3 &2.9\\ \hline 
 Been a victim at a previous organization &20.2 &18.8 &15.1 &8.7\\ \hline 
 Been a victim at multiple organizations &6.3 &7 &5.7 &4.3\\ \hline 
 Been a victim multiple times &20.2 &23.1 &10.8 &7.7\\ \hline 
 Been a witness &33.2 &32.8 &22.6 &14.9\\ \hline 
 Been  a biased actor  &7.1 &11.8 &1.9 &1.9\\ \hline 
    \end{tabular}
      \label{tab:bias-frequency}
\end{table}

\begin{table}

    \caption{Influence of the demographic factors on the likelihood of encountering bias. The asterisks (*) show the significance of the factor in bias experience and the values present the Odds Ratio.   For the categorical groups, '-' indicates, the reference category.  }
       \centering
    \begin{tabular}{|p{9em}|p{5em}|p{5em}|p{5em}|p{5em}|} \hline
\textbf{Attribute} & \textbf{CDB} &  \textbf{TSB } & \textbf{UE} & \textbf{IA} \\ \hhline{~----}

Model fit ($R^2$) & 0.416 $^{**}$ & 0.264 $^{**}$ & 0.19 $^{**}$& 0.346  $^{**}$ \\ \hline

  \multicolumn{5}{l}{\textbf{Age and Experience (Numerical)}} \\\hline
    
 Age & \textcolor{blue}{1.06 $^{*}$} & 1.01  & 1.03  & \textcolor{blue}{1.11$^{*}$} \\ \hline
 Dev. exp. & \textcolor{blue}{1.09 $^{*}$} & 1.04  & 1.02  & 1.01  \\ \hline
 Proj. exp. & \textcolor{blue}{0.88$^{*}$} & \textcolor{blue}{0.9$^{*}$} & 0.94   & 0.94   \\ \hline
 Org. size & \textcolor{blue}{0.85 $^{*}$} & 0.98  & 0.92   & 1.12   \\ \hline

 \multicolumn{5}{l}{\textbf{Compensation (Categorical)}}\\\hline
  Volunteer & -  & -   & -   & -   \\ \hline
 Paid & 0.76  & 1.33   & 1.66   & 1.75   \\ \hline

 \multicolumn{5}{l}{\textbf{Ethnicity (Categorical)}} \\\hline
 White & - & - & - & - \\ \hline
  Asian & 2.45   & 0.54   & 0.94   & 0.74   \\ \hline 
  African American & 2.39   & 1.03   & 1.73   & 1.82   \\ \hline
  Other Races & \textcolor{blue}{5.52$^{*}$} & 0.17   & 1.79   & \textcolor{blue}{11.51$^{**}$} \\ \hline
 Latino & \textcolor{blue}{6.88$^{**}$} & 1.03   & 0.36   & 3.98   \\ \hline

 \multicolumn{5}{l}{\textbf{Highest degree (Categorical)}}\\\hline
 High School &- &-&-&-\\\hline
College degree & 2.9   & 1.02   & 4.51   & 3.94  \\ \hline
 Postgrad & 1.26   & 0.62   & 2.99   & 1.72   \\ \hline

 \multicolumn{5}{l}{\textbf{Location (Categorical)}} \\\hline
 Location (NA) &-&-&-&-\\\hline
Location (EU) & \textcolor{blue}{0.41$^{*}$} & 0.57   & 1.01   & 1.72   \\ \hline
 Location (Rest)  & 0.79   & 2.03   & 1.54   & 1.99   \\ \hline

\multicolumn{5}{l}{\textbf{Gender (Categorical)}} \\\hline
   Man & - & - & - & - \\ \hline
  Woman & \textcolor{blue}{6.36$^{***}$} & \textcolor{blue}{4.67$^{***}$} & \textcolor{blue}{3.5$^{**}$} & 0.63  \\ \hline
  Non-binary& 1.41   & 1.15   & 2.05   & 0.48  \\ \hline
 
 \multicolumn{5}{p{12cm}}{ *** , **, and *  represent statistical significance at $p <$ 0.001, $p <$ 0.01, and $p <$ 0.05 respectively. No markers for $p$  indicate statistical insignificance. Values with statistical significance are colored in blue. }
\end{tabular}
\vspace{-20pt}

    \label{tab:demographic_regression}
    \end{table}

Table \ref{tab:bias-frequency} presents the percentage of respondents under different categories of bias. A total of 92 (36.4\%) of the 253 respondents acknowledged experiencing incidents involving CDB as a victim, witness, or biased actor. In addition, 110 (43.7\%) and 77 (30.7\%) respondents reported either directly or indirectly experiencing bias in TSB and UE.  Hence, one out of three respondents experienced incidents involving CDB, TSB, or UE. 
Regarding being a victim of CDB, 72 (28.5\%) respondents reported being one, while 84 (33.2\%) reported being a witness. A total of 18 (7.1\%) respondents acknowledged being biased actors of CDB. Among the CDB victims, 51 (71\% of the victims) reported multiple occurrences, and 16 (22\% of the victims) reported encounters at multiple organizations. 
For the TSB category, 84 (33.2\%) and 83 (32.8\%) acknowledged being victims and witnesses, respectively. Among the TSB victims, 59 (70\%) acknowledged being victims multiple times, and 17 (21\%)  reported encounters at multiple organizations.
While UE is less frequent than CDB and TSB, still 45 (17.9\%) and 57 (22.6\%) acknowledged being victims and witnesses, respectively. Among the UE victims, 27 (60\%) acknowledged being victims multiple times, and 15 (32\%)  reported encounters at multiple organizations. 
Finally, IA Although the ratio of respondents acknowledging IA is lower compared to the other three categories still, 46 (18.3\%) reported experiencing incidents with identity-demeaning jokes or insults, and 26 (10.1\%) reported being victims. Among the IA victims, 35 (76\%) encountered multiple times and 11 (42\%) encountered at multiple organizations.
We also noticed that the ratios of victims encountering CDB, TSB, UE, and IA at their current organization are 50\%, 49\%, 18\%, and 29\%, respectively. These numbers indicate that while occurrences of UE and IA are lower, more than two-thirds of their victims no longer work at incident organizations. 
 
\begin{boxedtext}
    \textbf{Takeaway 1:} CDB and TSB were the two most frequent forms of biases mentioned by software engineers from our sample. Being a victim was not a one-off occurrence for more than 70\% of the victims of each bias category.
\end{boxedtext}

\subsection{RQ1: Associations between demographic factors and being victims}
For this analysis, we have considered only the responses in which respondents acknowledged that they had been victims of an incident similar to the one described in a vignette. 
Table \ref{tab:demographic_regression} shows the associations between various demographic factors and the likelihood of being a victim. The pseudo-$R^2$ shows the performance of the four models. Since all models differ significantly from null models, they can reliably provide insights for this study. The numerical value next to each demographic attribute reflects the odds ratio (OR) in the logistic regression model for that variable. For numerical variables, OR above 1.0 indicates a higher likelihood of encountering a bias if that variable increases, and vice versa. For categorical variables, models assign a reference category for each demography.  For example, `White' is the reference category for the Ethnicity variable. OR $>1 $ for a non-reference category indicates a higher likelihood of encountering biases than the reference category, given all other variables are identical. Therefore, an OR value of 6.88 for the Latino group under CDB indicates that a `Latino' person is 6.88 times more likely to be the victim of a CDB than a `White' person, given all other demographics are identical between them. We summarize statistically significant demography associations for each bias category in the following.

\subsc{Career development bias (CDB)} Our analysis revealed significant associations between several demographic factors and the likelihood of experiencing CDB. First, the probability of becoming a victim of CDB increases with age and software development experience. This trend is unsurprising, as individuals with longer industry tenure are more likely to encounter CDB.
Second, we observed that the likelihood of experiencing CDB decreases as organization size grows. This finding suggests that CDB is more prevalent in smaller organizations, possibly due to limited opportunities for vertical growth. Third, longer project tenure is associated with a reduced likelihood of CDB. This aligns with prior research indicating that individuals facing workplace bias often leave their team or company \cite{griffiths2010disappearing}, resulting in shorter project tenures. Conversely, those with access to career growth opportunities tend to remain on projects longer. Our model also indicates that respondents from the European Union are significantly less likely to encounter CDB compared to their North American counterparts.
Furthermore, we found that individuals from underrepresented ethnic groups—such as Latino or ``Other races'' (e.g., American Indian, Alaska Native, Hawaiian, or Pacific Islander)—face odds of experiencing CDB more than five times higher than White individuals. Finally, this result corroborates prior studies showing that women are significantly more likely than men to be victims of CDB \cite{maji2020gendered,misa-dynamics-2021,margolis2002unlocking,o2021perceptions,elephant,tanwir2018breaking}.

\subsc{Task selection bias (TSB)} 
The TSB model suggests that as project experience increases, the odds of becoming a victim of TSB decrease. Therefore, newcomers are more frequently assigned tasks misaligned with their preferences, making them disproportionately susceptible to TSB. Our results also indicate that, holding all other demographic factors constant, a woman is 4.6 times more likely than a man to experience TSB.
These results are consistent with established patterns in the software development landscape, wherein women are more likely to be engaged in roles such as software quality assurance \cite{campero2021hiring}, front-end development \cite{breidenbach2021implicit}, and requirement elicitation~\cite{treude2023she}.

\subsc{Unwelcoming Environment (UE)}
The UE model suggests only one significant association. Even if all other demographics are identical, a woman is 3.5 times more likely to be a victim of UE than a man. This result also confirms recent studies~\cite{happe2021frustrations,chang2019brotopia}.

\subsc{Identity attack (IA)} The IA model suggests two demographics having significant associations. First, persons from ethnic minorities (i.e., Hawaiian/ Pacific Island/ American Indian) are 11.51 times more likely to be victims of jokes and insults targeting their racial identities than a White person. Second, we also notice that the odds of being a victim of IA increase with a person's age.

\begin{boxedtext}
   \textbf{ Takeaway 2: } Our results suggest that ethnic minorities are more likely to be targets of identity attacks (IA), while women, compared to men with otherwise similar demographics, are significantly more likely to experience CDB, TSB, and UE.
\end{boxedtext}

\subsection{RQ2: Consequences of biases}

We identified 62 codes on bias consequences and presented only the recurring ones in Table \ref{tab:qualitative_result}. Post hoc, we derived six higher-level categories for all the codes developed in the open-coding process. Additionally,  Table \ref{tab:qualitative_result} provides definitions for each category and lists the corresponding codes that fit into them.  We also fitted logistic regression models to identify if a person's demography is associated with how they react to a bias. 
We include qualitative discussions with excerpts for each category. The following sections detail our results.

\begin{table}

    \caption{Relationship among demographic factors and consequences of bias. The asterisks (*) show the significance of the demographic factor and the values present the Odds Ratio. For the categorical groups, '-' indicates, the reference category. }
    \centering
    \resizebox{0.9\linewidth}{!}{
    
\begin{tabular}{|p{9em}|p{5em}|p{5em}|p{5em}|p{5em}|}

\hline{~}
  &  \multicolumn{4}{l}{\textbf{Category of consequence on a bias victim} }  \vline\\ \hline

 \textbf{Attribute} & \textbf{Workplace} &  \textbf{Career} & \textbf{Emotion}  & \textbf{Psychology}  \\ \hline

Model fit ( $R^2$) & 0.295$^{**}$ & 0.32$^{***}$ & 0.286$^{***}$ & 0.377$^{***}$  \\\hline

\multicolumn{5}{l}{\textbf{Age and Experience (Numerical)}} \\ \hline
    
   Age & 1.01  & 1.03  & 1.03  & \textcolor{blue}{1.07 $^{*}$}  \\ \cline{1-5}
   Dev. exp. & 1.05  & \textcolor{blue}{1.08  $^{*}$}& 1.04   & 1.04   \\ \cline{1-5}
   Proj. exp. & 0.98   & 0.97  & 0.93  & 0.96    \\ \cline{1-5}
Org. size & \textcolor{blue}{0.81 $^{*}$} & 0.86  & 0.89  & 0.97   \\ \hline

\multicolumn{5}{l}{\textbf{Compensation (Categorical)}} \\ \hline
    Volunteer & - & -& -& - \\\cline{1-5}
     Paid & \textcolor{blue}{4.07 $^{*}$} & \textcolor{blue}{3.08 $^{*}$} & 1.7  & 1.41      \\ \hline

  \multicolumn{5}{l}{\textbf{Ethnicity (Categorical)}}\\\hline
   White & - & - & - & - \\\cline{1-5}
   Asian & 0.38   & 0.64   & 0.47   & \textcolor{blue}{0.26 $^{*}$}     \\ \cline{1-5}
    African American  & 1.97   & 0.67   & \textcolor{blue}{5.37  $^{*}$} & 1.74    \\ \cline{1-5}
    Other races & 1.1   & 0.67  & 2.57   & 0.37     \\ \cline{1-5}
    Latino & 1.76   & 2.17   & 2.42   & 1.15   \\ \hline 
    
 \multicolumn{5}{l}{\textbf{Highest Degree (Categorical)}}\\\hline
     High School &- &-&-&-\\\cline{1-5}
     College degree & 1.16   & 1.05  & 1.26   & 2.36     \\ \cline{1-5}
     Postgrad degree & 1  & 0.54   & 1.5   & 0.92    \\ \hline

   \multicolumn{5}{l}{\textbf{Location (Categorical)}}\\\hline
  Location (NA) &-&-&-&-\\\cline{1-5}
     Location(EU) & 0.75   & 0.41   & 0.73   & 0.86      \\ \cline{1-5}
  Location(Rest) & 1.47   & 0.57   & 1.54   & 2.33     \\ \hline

\multicolumn{5}{l}{\textbf{Gender (Categorical)}}\\\hline
     Man &-&-&-&-\\\cline{1-5}
      Woman & \textcolor{blue}{12.26  $^{**}$} & \textcolor{blue}{4.74  $^{**}$} & \textcolor{blue}{4.97  $^{***}$} & \textcolor{blue}{5.44  $^{***}$}    \\ \cline{1-5}
     Non-Binary & 2.25   & 0.42   & 2.13   & 0  \\\hline

\multicolumn{5}{p{12cm}}{ *** , **, and *  represent statistical significance at $p <$ 0.001, $p <$ 0.01, and $p <$ 0.05 respectively. No markers for $p$  indicate statistical insignificance. Values with statistical significance are colored in blue. }
\end{tabular}
 
    }   
    \label{tab:consequence_regression}
   \end{table}

\begin{table}
    \caption{Higher level categories of bias consequences and their definitions}
    \centering    
    \resizebox{\linewidth}{!}{
\begin{tabular}{p{2cm}p{4cm}p{7cm}}
\rowcolor[HTML]{9B9B9B} 
Category   & Definition  & Codes  \\
\rowcolor[HTML]{EFEFEF} 
Workplace dissatisfaction \& tension (Workplace)& 
Comprises feelings of discontentment with one's current job or organization, as well as negative dynamics in the interaction with colleagues 
&  Intended to quit, lost interest in job, lost respect for manager/company/colleagues, frustrated in the company, strained relationship with colleagues, stopped/limited interaction with the offender, affected team bonding, talked to offender/manager, less engaged with the offender/other people\\

Career trajectory (Career)& Includes concrete consequences in career or current job  & Left job/team, income downfall, performed extra work, changed career track, volunteered/asked for challenging tasks, tried harder to prove themselves \\%

\rowcolor[HTML]{EFEFEF} 

Emotional \& general discontentment (Emotion)& Involves short-term emotional effects  & Upset, uncomfortable, angry, bitter feeling, underappreciated, helpless, isolated, shocked, puzzled\\


Effects on psychological well-being (Psychology)& Includes the challenges people encounter psychologically for long-term which hamper their overall well-being and self-esteem& 
Demoralized, lost confidence, self-doubt, depressed, having imposter syndrome, felt vulnerable, feeling disrespected, fear of harassment, frightened\\


\rowcolor[HTML]{EFEFEF} 

Disciplinary actions (disciplinary) & Encompasses the disciplinary actions or measures implemented by management in response to workplace misconduct  
& The offender was transferred/fired, the offended person was fired , the offended person complained , management took action, the manager warned the offender, the offender was transferred or fired, the bias receiver/ anti-harassment team took action \\

Considered bias as norm (norm) & Acknowledging the existence of bias and accepting it as a norm  
& Considered bias as a norm,  moved on \\

\hline

\end{tabular}       

}     
    \label{tab:qualitative_result}
\end{table}

\subsc{Associations between categories of consequences and bias types}  Figure \ref{fig:result_sankeymatic}, shows the relationships between the types of biases and the categories of influences reported by the respondents. The effect on psychological well-being was the most frequently mentioned category by the respondents (111 individuals). Among them, 52 persons reported experiencing this effect due to CDB, 25 for TSB, 23 for UE, and 11 for IA. Emotional and general discontentment in life were mentioned by 42, 30, 15, and 12 participants, respectively, for CDB, TSB, UE, and IA. Similarly, the effect on one's career or current job was mentioned by 37, 24, 14, and 6 participants, respectively.
Five respondents indicated biased actors facing workplace disciplinary actions for IA and UE, which is only 8\% of the cases reported in our survey. 
Finally, a few victims of CDB, TSB, and UE considered such types of bias as common or norm in the workplace.

\subsc{Associations between demographic factors and bias consequences} 
Table \ref{tab:consequence_regression} presents the associations between demographic factors and the various categories of influences resulting from the mentioned biases. While we trained six models, one for each category of consequences, two models were not better than a null model, as suggested by the log-likelihood test, and therefore are not presented. The results indicate that women have more than 12 times higher odds than men of being dissatisfied with the workplace due to being a biased victim. Similarly,  biases are significantly more likely to cause three other reactions from women than from men, which include altered career trajectory, instigating short-term emotional responses, and even causing long-term psychological effects. Our results also indicate that as organizational size increases, victims are less likely to experience workplace tension and dissatisfaction, possibly due to increased flexibility in changing teams.
Moreover, paid professionals are more likely to experience workplace dissatisfaction and career impediments due to biases. As work experience increases, the likelihood of being dissatisfied with a career trajectory due to biases increases. Our results also suggest that African-American participants are more likely to be affected emotionally, while Asian participants are less likely to be affected psychologically. Finally, our model suggests that vulnerability to long-term psychological consequences due to biases increases with age.

\begin{figure}
	  \includegraphics[width=0.7\linewidth]{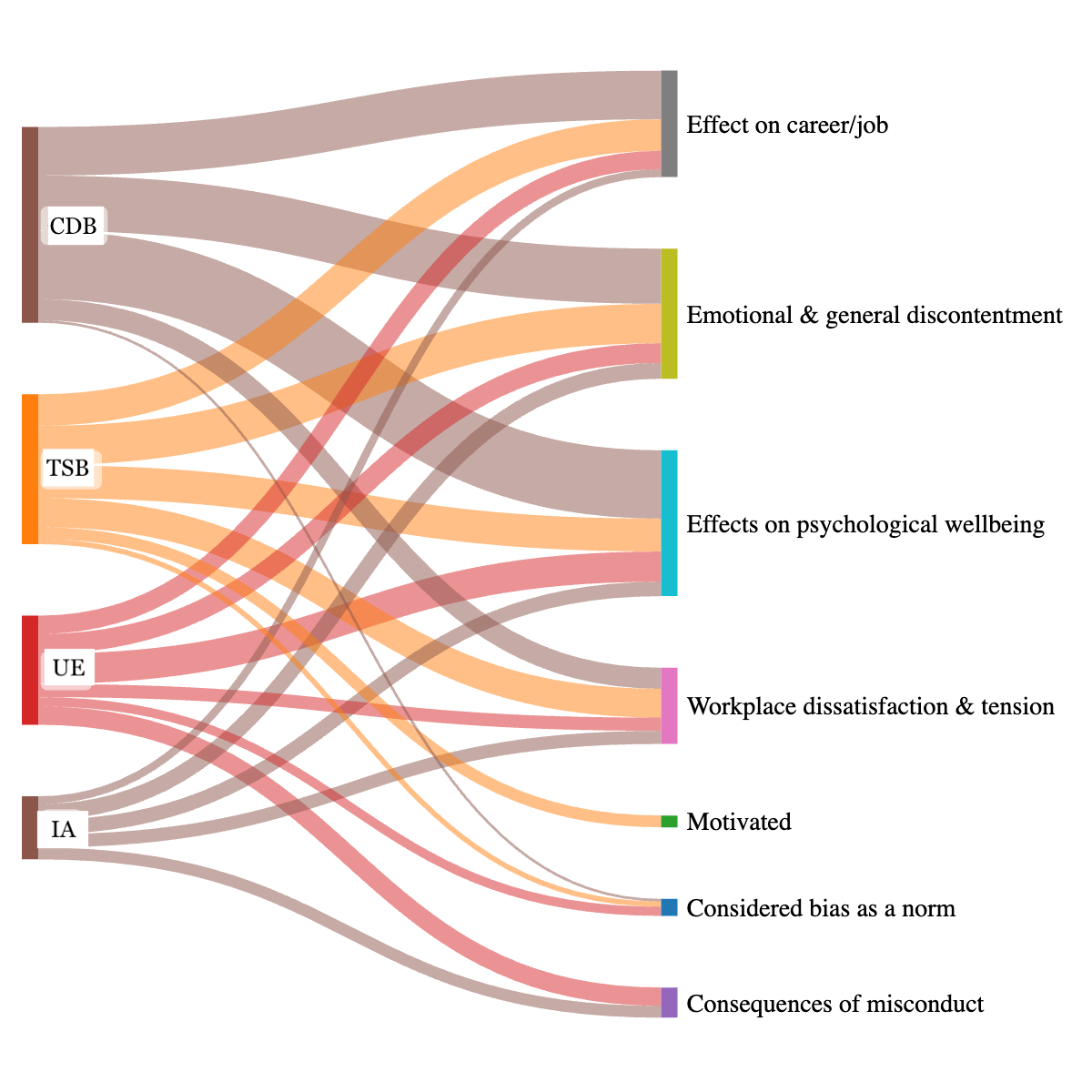}  
\caption{The associations between bias types and categories of bias consequences}
\label{fig:result_sankeymatic}	
\end{figure}

\subsc{Consequences of CDB} People become frustrated and demotivated due to experiencing such biases in the workplace. Many lose interest in jobs, engage less in the workplace, and eventually leave.
The most frequently cited outcome of CDB was either changing teams or departing from the job, as mentioned by 27 respondents, indicating that bias victims often chose to leave their positions. One of them shared: \excerpt{I left the first company. I did not want to work for that company anymore.} 
A large number of respondents also mentioned bitter feelings that impacted their psychological well-being. 
\excerpt{I felt confused, doubted myself more and felt resentful. I did not feel I was in control or could trust my instincts. The problem is you never exactly know why it happens, if it really was gender stereotyping or if you were just `not good enough' or if it was something else.} 

Eight respondents said they changed their team or started looking for another job. After being the recipient of unfair evaluation, they lose interest in the job and intended to quit. \excerpt{I got it fixed (they did not want to give a rectification) but the next assessment was a "good" again. But I felt no longer a connection with the company. } 
Two respondents mentioned their evaluation deteriorated after being mothers, and promotion was delayed. Among them, one respondent shared her experience as:\excerpt{About three years ago, because I was a new mother, the company leadership decided that I didn't have enough energy to finish my work project and handed it over to someone else without any discussion, ... I feel very helpless.}
Moreover, such a type of bias gives rise to anger, confusion, self-doubt, and feeling of unfairness. \excerpt{I felt confused, doubted myself more, and felt resentful. I did not feel I was in control or could trust my instincts.} 



\subsc{Consequences of Task selection bias (TSB)} Many respondents reported that bias victims experienced emotional distress, feeling upset, angry, unappreciated, and singled out due to encountering such bias. This bias also led to strained relationships with colleagues and caused frustration within the company, resulting in performance deterioration. Some respondents expressed their intention to quit the job (4 participants) or had already left the job (8 individuals) due to TSB. One respondent shared the influence of such bias as \excerpt{A couple of people have left the org because of how they’ve been treated and how people have not believed in their skills.}
However, eleven respondents shared that rather than giving up, they were motivated to work harder to overcome the biased perceptions they faced. For example, one respondent shared, \excerpt{I felt determined to keep working harder and ask for harder tickets.} 
However, many developers considered such bias as a norm and started looking for other job opportunities. One respondent shared as, \excerpt{Participants took that normally. However, they are planning to shift company. That takes effect after 3-4 months.} 

Some victims of bias felt disgusted and believed they were hindered by a "glass ceiling" that prevented them from advancing--  \excerpt{Offenders kept their upward trajectory of the best projects and promotions; offended had speed knots from bumping their heads on the glass ceiling.} 
After encountering the bias, many receivers start acting defensively and work harder to prove themselves again and again. It also influences the attitude of the biased receiver to a great extent. One witness of such incident shared, \excerpt{My coworker lost her trust in anyone in the company, she acts defensive and always wants to prove herself.} TSBs not only influence the receiver personally but also affect the relationship among co-workers. \excerpt{There was a slight disconnect of unity felt across the team.} 

\subsc{ Consequences of Unwelcoming Environment (UE)} Many respondents were hurt by UA and also surprised by in-actions from their colleagues as if those are norms-- \excerpt{Most people seemed to avoid the subject completely. As if it hadn't happened. Many pretend to not know or remember(?). The thing that surprised me the most was how people didn't try to help the person being abused/harassed.} Hence, some victims had to leave their jobs, even if they had to sacrifice career prospects. \excerpt{I left my job and had to rebuild my career in a different area. I had to take junior positions in the new field and lost salary as a consequence.}
 
Moreover, such types of attitudes at work make people insecure and result in strained relationships. One respondent shared that,\excerpt{I blocked that person and started to feel so insecured.} 
Getting asked personal questions also make people afraid to interact with people. Another respondent shared,\excerpt{...but he always asked me some personal questions, which had little to do with work, this kind of thing makes me afraid to talk to people. I think it's covert harassment.}
Many women also do not have an appropriate place to call for help in these types of scenarios. They feel embarrassed and console themselves as if nothing bad incident occurred. One respondent shared the influence as follows: \excerpt{I could only politely refuse. I can't turn to anyone else for help with this kind of thing. After all, nothing bad has happened. I feel so embarrassed.} 

\subsc{ Consequences of Identity attack (IA)} Many respondents said they had encountered identity attacks. People feel helpless, insulted, confused, and uncomfortable when they fall victim to IAs. 
\excerpt{My boss wanted to send out an e-mail campaign to promote our business to all our business’ clients, which would be making a joke of my sexual orientation. He asked me what he thought and I dared not say no. I did refuse to participate later. But it made me feel very disturbed and belittled.} 
As a result of such emotional influence, people stop sharing their personal information at work and limit their engagement with colleagues, which eventually breaks the team spirit. When the victims protest to management, sometimes they are urged to ``grow a thicker skin" and that their complaints would not be taken seriously. One respondent said, \excerpt{Once my teammates discovered I am gay, the joking and teasing began. It was routinely dismissed by management and HR as “boys being boys.” I was told to lighten up and that they were only jokes and I needed to grow thicker skin.} 

\begin{boxedtext}
    \textbf{Takeaway 3:}  Biases in the workplace lead to psychological distress, and job attrition, also create self-reinforcing cycles of exclusion and self-doubt where victims disengage or leave, further strengthening in-group dominance. Also, lack of institutional support and accepting the bias as a workplace culture suppress intervention efforts and support systemic inequities. Vulnerability to long-term psychological consequences due to biases increases with age. 
\end{boxedtext}


\subsection{RQ3: Motivation behind biased decision}
A total of 33 responses acknowledged being biased actors. 
Table \ref{tab:motivation_result} shows the biased actors and their motives.

\begin{table}
    \centering
    \caption{Motivation of the biased actors}
    \label{tab:motivation_result}

\resizebox{\linewidth}{!}{
\begin{tabular}{|p{5cm}|p{5cm}|R{1.9cm}|}
\rowcolor[HTML]{9B9B9B} 
Bias type   (\# of respondents)                   & Code     & \% of resp. (\#)     \\

\multirow{4}{5cm}{Career development bias (7)}& Motivated by coworkers & 14.29\% (1) \\ \cline{2-3}
    & Decision of upper management & 42.86\% (3)\\ \cline{2-3}
    & Incompetent women & 28.57\% (2)\\ \cline{2-3} 
    & Preference to work with men & 42.86\% (3) \\ 
\hline

\multirow{4}{5cm}{Task selection bias (19)}& Unconsciously stereotyped & 47.37\% (9)\\ \cline{2-3}
    & Benevolent sexism & 21.05\% (4)\\ \cline{2-3}
    & Women's lack of confidence & 21.05\% (4)\\ \cline{2-3}
    & To motivate women more & 5.26\% (1)\\
   
\hline

\multirow{1}{5cm}{Unwanted attention (3)}& Not sure/ thought of not a big deal & 100\% (3) \\ 
\hline

\multirow{2}{5cm}{Identity attack (4)}& Passively supporting in group & 50\% (2)\\ \cline{2-3}
    & Casual joke & 50\% (2)\\ 
\hline
\end{tabular}       

}  
    \end{table}
\subsc{Motivation behind CDB.}  One biased actor mentioned assuming that women are less competent, and his preference to work with men, \excerpt{I assumed the woman was less competent. I did not make such a big decision though. I just always preferred to work with men. It was more a sum of smaller decisions.} Others shared that they were motivated by their co-workers to make biased decisions, and they felt compelled to follow instructions from upper management. One respondent mentioned choosing a male leader for the long-term within the company, assuming that women would not be able to serve in that position for an extended period.

\subsc{Motivation behind TSB.} For the task selection bias, nine respondents shared of having unconscious stereotypes, and they blamed the workplace's stereotypical culture for these events.
\excerpt{I am assuming that since I am a product of this environment, it is possible I was also mean that way sometimes and didn't even realize that I was doing it.} Sometimes, such biased incidents occur as a result of benevolent sexism. The discriminators offer extra help to the victims, which might influence the bias receiver negatively. For example, one respondent said, \excerpt{I worked with a female junior developer and offered to do a lot of her workload in order to make her look better, and yet would still give her all the credit. } 
One respondent shared that a biased attitude can motivate the victims, and one person noted that he showed bias from prior experience, \excerpt{The other person involved was a female and she felt bad that I didn't think she was up to the task. My motivation was that in the past, whenever we were being assigned projects and tough tasks were being given to her, most of times she didn't complete them or she got stocked.} Moreover, four respondents considered lack of confidence from women's side in assigning them low-impact assignments.
\excerpt{I underestimated the technical skills of a junior woman because of her speech patterns: she tended to hedge and express uncertainty in situations where her male peers tended not to. It took me a long time to notice that in fact her work was as good or better than theirs, and in fact her expressions of uncertainty were well-calibrated, whereas the confidence of her male counterparts was unjustified. I worked out what was happening eventually, it just took me longer than it should have.}

\subsc{Motivation behind UE.} We found that while women might feel uncomfortable, men perceive paying unwanted attention differently and do not consider it as harmful. One person shared that, \excerpt{My motive was to be close to her. I had no idea I was having such an impact on her.} Some respondents indicated they were unsure if their behavior was unwelcome since they never got reported to HR.
\excerpt{Again, that maybe is more of an "I'm not sure I never did?". Never got anything escalated to HR, and never got the feeling of being inappropriate, but maybe I missed signals and was actually being creepy.}

\subsc{Motivation behind IA.} Two respondents shared that they were not actively involved in such incidents but passively supported in a group.\excerpt{Probably laughed along at one point when someone else made fun of being gay.}. Also, two other respondents shared that they sometimes made casual jokes and did not think that might be harmful to the victims. \excerpt{At that time only wanted to make a joke, and did not expect the consequences.}

\begin{boxedtext}
    \textbf{Takeaway 4:} Biases in the workplace often stem from unconscious stereotypes, social pressure, and workplace culture. Our respondents who had biases against other software developers would rather be working with men and showed benevolent sexism against women. Some were unaware of the harm they caused, while others passively supported biased behavior.
\end{boxedtext}

\section{Discussion}
\label{sec:discussion}

\vspace{2pt}
{\noindent \textbf{Extending SIT through an intersectional lens:} Our findings corroborate Social Identity Theory (SIT), confirming that in-group favoritism and out-group bias are prevalent within software engineering workplaces. Furthermore, we refine SIT by applying an intersectional lens, demonstrating that a developer's experience of bias is modulated by compounding identity factors, including gender, race, experience, age, and organizational size. For instance, our data suggests that Latina women and older women are particularly vulnerable to CDB. Additionally, comparatively older individuals from minoritized groups—specifically those who do not identify as White, Asian, Latino, or Black—are statistically more likely to experience IA.}

\vspace{2pt}
\noindent \textbf{Groups susceptible to double jeopardy:} Results from Table \ref{tab:demographic_regression} indicate that specific subgroups are disproportionately vulnerable to biases, supporting the concept of `double jeopardy'~\cite{williams2014double}. The analysis reveals that age and professional experience are significantly correlated with career development bias, a trend that is particularly pronounced for women; consequently, older women face heightened susceptibility to CDBs. Additionally, while Latina women demonstrated notably higher odds of encountering CDBs, women located in Europe appeared less likely to report such incidents. Furthermore, individuals falling outside the major racial categories (White, Asian, Latino, or Black) are more susceptible to identity attacks. Given that factors beyond gender—such as age, race, and geography—significantly influence the biases encountered by software engineers, it is imperative to examine gender bias in this domain through a rigorous intersectional framework~\cite{crenshaw2013mapping}.

\vspace{2pt}
\noindent\textbf{Recommendations for minoritized professionals:}
First, as more than 70\% of victims reported encountering biases on multiple occasions, these incidents should not be viewed as isolated anomalies but as patterns requiring mitigation strategies.
Second, qualitative data revealed that some biased actors remained unaware that their behavior was offensive because it went unreported, effectively interpreting silence as implicit acceptance~\cite{hastie2021unwelcome}. Therefore, where safe to do so, individuals may consider addressing issues directly with biased actors, supervisors, or HR by openly communicating concerns and documenting specific examples to substantiate claims.
Third, our results indicate that CDBs are more prevalent in smaller organizations. Consequently, early-career professionals from marginalized groups might consider prioritizing larger organizations, which often possess more robust structural protections and resources, thereby potentially reducing the risk of biases that stifle career progression.
Finally, professionals should familiarize themselves with relevant workplace policies and employment laws. A strong understanding of one's rights can empower individuals to navigate bias incidents more effectively. Furthermore, as individuals from underrepresented ethnicities may be the sole representative of their group within an organization, internal support may be scarce. In such cases, cultivating a professional network outside the organization and connecting with peers of similar backgrounds can provide essential solidarity and advice.

\vspace{2pt}
\noindent\textbf{Recommendations for managers and authorities:}
Organizational policies, leadership practices, and cultural norms fundamentally influence the prevalence of bias. Consequently, senior leadership plays a critical role in either perpetuating or mitigating these issues.
First, leadership must institutionalize diversity and inclusion awareness. Some biased actors admit to making decisions based on unconscious prejudice. According to SIT, such biases often stem from stereotypical beliefs regarding the roles of specific demographics~\cite{treude2023she,campero2021hiring} or affinity bias—the preference for working with peers of similar backgrounds. These biases distort decision-making and reinforce barriers for minorities. To utilize the `contact hypothesis,' authorities should implement structured mentorship programs that connect out-group members with in-group members. This exposure can help in-group individuals recognize the unique challenges faced by their out-group counterparts, thereby reducing implicit bias and fostering positive workplace interactions~\cite{why_diversity_programs_fail}. Additionally, mandatory training on unconscious bias can equip team members with the tools to recognize and address their own cognitive shortcuts.
Second, our results identify specific populations—including women, individuals from marginalized ethnicities, and older professionals—who are disproportionately vulnerable to bias and its severe consequences. Project managers must actively monitor the well-being and inclusion of these subordinates. Managers should also participate directly in diversity initiatives, such as targeted recruitment programs; engaging in these efforts can help dismantle their own biased perceptions over time~\cite{why_diversity_programs_fail}.
Third, given that more than two-thirds of IA and Unfair Evaluation (UE) victims leave the organization where the incident occurred—compared to 50\% of CDB and Technical Skill Bias (TSB) victims—proactive mitigation is essential for retention.
Finally, as many victims leave organizations due to a lack of managerial support, companies must enforce social accountability. Beyond adopting inclusive policies, organizations should consider transparency measures, such as publicly sharing aggregated team performance ratings and pay raises by gender and race. This transparency highlights areas of the organization that may be demonstrating favoritism, holding managers accountable for equitable treatment~\cite{why_diversity_programs_fail}.

\vspace{2pt}
\noindent \textbf{Recommendations for researchers:}
First, our results demonstrate that susceptibility to bias is significantly associated with a complex interplay of demographic factors, including age, gender, ethnicity, location, and organization type. Therefore, a `one-size-fits-all' approach is inadequate for promoting equitable participation. Universal strategies may fail to address the nuance of marginalized groups' experiences. Consequently, further studies are required to disentangle these associations and design targeted interventions. Second, understanding the underlying motives of biased actors is crucial for developing effective mitigation strategies. Due to sample size limitations, this study could not exhaustively investigate these motives or their intersections with demography. Future work should prioritize this area. Finally, while our analysis suggests `double jeopardy' for specific subgroups, quantitative validation was limited by sample size. A promising avenue for future research is to conduct in-depth qualitative interviews with participants from these intersectional groups to provide a richer, more nuanced understanding of their lived experiences with bias.

\section{Threats to Validity}
\label{sec:threats}
 \textbf{Internal validity:} Due to survey design constraints, we focused on four aspects of biases likely to arise from social identity. However, we recognize that additional types of bias could affect women and other minorities, and all bias types warrant in-depth investigations to promote diverse computing organizations.
 Since we shared the survey URL through multiple channels, which include social media, developer-oriented groups, and mailing lists, estimating the response rate is infeasible.
 Although our results may be susceptible to \textit{non-response bias}~\cite{olson2006survey}, a total of 253 quality-verified responses provide us with a valuable and comprehensive dataset to identify key insights relevant to the goal of this study. We also acknowledge \textit{self-selection bias}~\cite{olson2006survey} that involves people who care about promoting diversity and inclusion being more likely to respond to our survey, as our solicitation includes those words. 

 Since Bias is a sensitive and socially charged topic, even with anonymous data collection, participants may have exhibited social desirability bias. Victims might hesitate to report incidents that they fear could be interpreted as personal failures or "complaining," while witnesses might underreport their peers' behavior to protect group cohesion. Conversely, the vignettes might have lowered the threshold for what participants consider "bias," leading to an over-reporting of minor interpersonal conflicts as systemic bias. To address this, ensured strict anonymity to encourage honest and uninhibited reporting.
 Moreover, to ensure reliability, two researchers independently coded the qualitative data using an inductive coding approach. The annotators differed in both gender and age group to enhance diversity in interpretation.

 \vspace{2pt}
 \noindent \textbf{Construct validity:} 
A primary threat to construct validity in this study arises from the use of vignettes. While vignettes were employed to aid recall and clarify the definition of bias, they carry the risk of priming participants. By presenting specific scenarios of bias (e.g., CDB or IA), we may have inadvertently focused participants' attention on incidents similar to those described, potentially leading to the underreporting of other, less prototypical forms of bias. Furthermore, the vignettes represent simplified, archetypal interactions that may lack the nuance, history, and ambiguity of real-world workplace dynamics. To mitigate this, we emphasized in the survey instructions that the vignettes were illustrative examples rather than an exhaustive list, encouraging participants to report any incident they perceived as biased regardless of its similarity to the provided scenarios. 

The retrospective nature of the survey, combined with vignette stimulation, introduces the potential for recall bias. Participants were asked to remember past events, which can be subject to memory decay or distortion over time. There is also a risk that reading a vignette could trigger false recognition or reconstructive memory errors, where participants might subconsciously align their vague memories with the clear narrative of the vignette. We attempted to minimize this by asking follow-up questions regarding specific details of the incident (e.g., the role of the biased actor, the outcome) to ensure the reported events were grounded in specific, retrieveable episodes rather than general feelings of unfairness.
 
Another concern was that respondents might have difficulty understanding the questions or misinterpret the scenarios. To mitigate this, we sought expert input and incorporated their suggestions in the survey design. Additionally, we conducted a pilot study to ensure the questions were clear and appropriate for the respondents. To avoid any biased responses, we refrained from explicitly using the term `bias' in the scenarios, questions, or invitations. This approach aimed to prevent participants from being triggered and encouraged them to answer in an unbiased manner. Furthermore, we provided clear instructions at the beginning of the survey, emphasizing that respondents should base their answers on their personal experiences. 
During the analysis, we employed iterative coding and thematic analysis, with multiple researchers reviewing qualitative responses to ensure consistent interpretation and minimize subjectivity. We also employed methodological triangulation by combining survey-based quantitative responses with open-ended qualitative insights.

  \vspace{2pt}
\noindent \textbf{External validity:} 
While the vignettes were designed to be realistic, they may not perfectly capture the cultural nuances of bias across all geographic locations represented in our global sample. A scenario perceived as an IA in a Western workplace might be interpreted differently in a non-Western context due to differing communication norms and power distance indices. Consequently, the interpretation of the vignettes may vary across demographic groups, potentially affecting the consistency of the data. We acknowledge this limitation and suggest that future work employs locally adapted vignettes to validate these findings across distinct cultural contexts.

Moreover, the analysis of motivations for biased behavior is constrained by self-reporting bias. Participants are unlikely to admit to actions that could be perceived as unethical or offensive, even in an anonymous survey. This reluctance likely contributed to the small number of responses from this group. While understandable, this low response rate highlights a critical challenge. Further research is needed to better understand the motivations behind discriminatory behavior, enabling the development of more effective interventions.

Finally, our findings reflect the perspectives of our survey participants. Although the sample was demographically diverse, the results may not be generalizable to all software development organizations.

\section{Conclusion}\label{conclusion}

This study offers a comprehensive examination of the factors that contribute to biased attitudes and decisions in computing. Our findings confirm that bias is a multifaceted problem. We show that beyond gender and race, factors such as age, years of experience, organization size, and geographic location are also significant predictors of bias victimization. This complexity reveals that a one-size-fits-all approach to bias mitigation is insufficient.

The critical insights from this research, which examines how different biases impact workplace satisfaction, well-being, and career trajectories, can help organizations develop more targeted and effective interventions. A comprehensive strategy should encompass raising awareness of these nuanced biases, implementing inclusive policies, and providing robust support for individuals affected by them. Without such informed efforts, the full potential of underrepresented groups will go untapped, resulting in missed opportunities for innovation and progress across the field. 

\subsection*{DATA AVAILABILITY}
\label{Data_availability}

The dataset, source code, and experimental results have been provided as supplementary materials are publicly available at: \url{https://doi.org/10.5281/zenodo.18408986}.

\bibliographystyle{ACM-Reference-Format}  
\bibliography{bibliography} 

\end{document}